\documentclass[showpacs,floatfix,citeautoscript,twocolumn,prb,superscriptaddress]{revtex4-1}

\usepackage{graphicx,xcolor}
\usepackage{amsmath}
\usepackage{comment}
\usepackage{textcomp}
\usepackage{inputenc}

\newcommand*{\aas}{Ag$_{3}$AuSe$_{2}$}
\newcommand*{\aat}{Ag$_{3}$AuTe$_{2}$}
\newcommand{\agse}{Ag$_{2}$Se}

\begin{document}

\title{Transport and optical properties of the chiral semiconductor Ag$_3$AuSe$_2$}

\author{Juyeon Won}
\affiliation{Department of Materials Science and Engineering
and Materials Research Laboratory, University of Illinois at Urbana-Champaign, Urbana, IL 61801, USA}

\author{Soyeun Kim} 
\affiliation{Department of Physics, University of Illinois at Urbana-Champaign, Urbana, 61801 IL, USA}
\affiliation{Materials Research Laboratory, University of Illinois at Urbana-Champaign, Urbana, 61801 IL, USA}

\author{Martin Gutierrez-Amigo}
\affiliation{Department of Physics, University of the Basque Country (UPV/EHU), Apartado 644, 48080 Bilbao, Spain}
\affiliation{Centro de F'{i}sica de Materiales (CSIC-UPV/EHU), Manuel de Lardizabal Pasealekua 5, 20018 Donostia/San Sebasti\'{a}n, Spain}

\author{Simon Bettler} 
\affiliation{Department of Physics, University of Illinois at Urbana-Champaign, Urbana, 61801 IL, USA}
\affiliation{Materials Research Laboratory, University of Illinois at Urbana-Champaign, Urbana, 61801 IL, USA}

% SNU person (students)
\author{Bumjoo Lee} 
% kimphysics@snu.ac.kr
\affiliation{Center for Correlated Electron Systems, Institute for Basic Science, Seoul 08826, Republic of Korea}
\affiliation{Department of Physics and Astronomy, Seoul National University, Seoul 08826, Republic of Korea}

\author{Jaeseok Son} 
% zangcro@snu.ac.kr
\affiliation{Center for Correlated Electron Systems, Institute for Basic Science, Seoul 08826, Republic of Korea}
\affiliation{Department of Physics and Astronomy, Seoul National University, Seoul 08826, Republic of Korea}

% SNU person (professor)
\author{Tae Won Noh} 
% twnoh@snu.ac.kr
\affiliation{Center for Correlated Electron Systems, Institute for Basic Science, Seoul 08826, Republic of Korea}
\affiliation{Department of Physics and Astronomy, Seoul National University, Seoul 08826, Republic of Korea}

\author{Ion Errea}
\affiliation{Donostia International Physics Center, P. Manuel de Lardizabal 4, 20018 Donostia-San Sebastian, Spain}
\affiliation{Centro de F'{i}sica de Materiales (CSIC-UPV/EHU), Manuel de Lardizabal Pasealekua 5, 20018 Donostia/San Sebasti\'{a}n, Spain}
\affiliation{Fisika Aplikatua 1 Saila, Gipuzkoako Ingeniaritza Eskola, University of the Basque Country (UPV/EHU), Europa Plaza 1, 20018 Donostia/San Sebasti\'an, Spain}
\author{Maia G. Vergniory}
\affiliation{Donostia International Physics Center, P. Manuel de Lardizabal 4, 20018 Donostia-San Sebastian, Spain}
\affiliation{Max Planck Institute for Chemical Physics of Solids , 01187 Dresden, Germany}

\author{Peter Abbamonte} 
\affiliation{Department of Physics, University of Illinois at Urbana-Champaign, Urbana, 61801 IL, USA}
\affiliation{Materials Research Laboratory, University of Illinois at Urbana-Champaign, Urbana, 61801 IL, USA}

\author{Fahad Mahmood} 
%\email{fahad@illinois.edu}
\affiliation{Department of Physics, University of Illinois at Urbana-Champaign, Urbana, 61801 IL, USA}
\affiliation{Materials Research Laboratory, University of Illinois at Urbana-Champaign, Urbana, 61801 IL, USA}

\author{Daniel P. Shoemaker}\email{dpshoema@illinois.edu}
\affiliation{Department of Materials Science and Engineering
and Materials Research Laboratory, University of Illinois at Urbana-Champaign, Urbana, IL 61801, USA}

%\date{\today}

\begin{abstract}
Previous band structure calculations predicted \aas\ to be a semiconductor with a band gap of approximately 1~eV. Here, we report single crystal growth of \aas\ and its transport and optical properties. Single crystals of \aas\ were synthesized by slow-cooling from the melt, and grain sizes were confirmed to be greater than 2~mm using electron backscatter diffraction. Optical and transport measurements reveal that \aas\ is a highly resistive semiconductor with a band gap of and activation energy around 0.3~eV. Our first-principles calculations show that the experimentally-determined band gap lies between the predicted band gaps from GGA and hybrid functionals. We predict band inversion to be possible by applying tensile strain. The sensitivity of the gap to Ag/Au ordering, chemical substitution, and heat treatment merit further investigation. 
\end{abstract}

\maketitle

\section{Introduction} 

\aas\ was first reported reported in 1971 as the mineral fischesserite from P\v{r}edbo\v{r}ice, Czech Republic, by Johan, et al.\cite{johan_fischesserite_1971} 
It was the first gold-containing selenide reported, and its crystal structure was identified as isostructural to the chiral mineral petzite Ag$_3$AuTe$_2$ in space group $I4_132$. Moderate reflectivity in the visible range and a Moh's hardness of two were corroborated on another natural sample in 2004.\cite{bindi_structural_2004} 
Initial work on the silver-gold chalcogenides used differential thermal analysis (DTA) to show that they undergo order-disorder transitions to body-centered cubic Ag$_2$S phases upon heating, with \aas\ transforming at 270$^\circ$C.\cite{tavernier_ber_1967}
The tellurite \aat\ undergoes two transitions at 220$^\circ$C and 320$^\circ$C, with the intermediate crystal structure unknown, even though high-temperature X-ray diffraction (XRD) data were collected.\cite{smit_phase_1970} The sulfide uytenbogaardtite, on the other hand, has a more complex distorted tetragonal $P4_122$ structure at room temperature\cite{barton_uytenbogaardtite_1978}  and a transition at 185$^\circ$C.\cite{tavernier_ber_1967}

The first  studies on synthetic fischesserite by Smit, et al.\cite{smit_phase_1970} focused on confirming the previously-reported phase transitions by DTA and XRD.

Wiegers\cite{wiegers_electronic_1976} investigated the electronic and ionic conductivity of \aas\, with a claimed band gap from diffuse reflectance spectroscopy of 0.9~eV, but the samples were processed at 450$^\circ$C and no evidence of their purity (XRD or otherwise) was provided. 
%Wiegers seebeck says p-type
The remaining work on synthetic \aas\ include a M\"{o}ssbauer study showing that the ordered silver-gold chalcogenides contain monovalent Au$^+$ and stronger bonding  than the binary chalcogenides,\cite{wagner_mossbauer_1992} and electromotive force measurements of thermodynamic equations of state that confirmed the stability of these compounds but did not precisely agree with the order-disorder transition temperatures. \cite{osadchii_system_2007,feng_thermodynamic_2014}

\aas\ and \aat\ are chiral and isostructural at room temperature with the space group of $I4_132$. The chirality and possible narrow band gap (or metallicity) of these materials gives rise to possible non-reciprocal and topological phenomena. 
In this report, we investigate the room-temperature phase of \aas. 

More recently, the chirality and possible narrow band gap (or possibility of band inversion and topological phenomena) has led renewed computational focus on \aas. Faizan used density functional theory (DFT) to evaluate the promise of \aas\ for thermoelectric applications, and found the band gap to be about 1.0~eV using the generalized gradient approximation (GGA).\cite{faizan_elastic_2016,faizan_carrier_2017} 
However, published resistivity measurements (with the aforementioned question of sample purity)\cite{wiegers_electronic_1976} gave a lower activation energy of around 0.5~eV. Calculations using the local density approximation (LDA), by Fang, et al. are expected to underestimate the band gap and found a gap of about 0.2 eV. \cite{fang_ab_2002}
Bradlyn and collaborators\cite{Bradlynaaf5037} predicted it to host multifold topological crossings. Since 2016 multifold crossing in chiral topological semimetals have attracted great attention owing to their large Chern numbers \cite{Bradlynaaf5037,PhysRevLett.119.206401}, long Fermi arcs \cite{schroter2019chiral,schroter2020observation,sanchez2019} and optical quantized responses in terms of Chern numbers.\cite{flicker2018chiral} 
Better understanding the band separation in \aas\ is required to determine whether closing this narrow gap could lead to topological insulating behavior was examined by Sanchez-Martinez\cite{sanchez-martinez_spectral_2019}.
DFT-GGA calculations showed that decrease in lattice parameters result in band gap widening. In addition, they found that compression of 2\% yields a meV-range band gap, which may be suitable for dark matter detector applications.
All previous computational studies have pointed toward a direct band gap, but no single crystal measurements have been made, nor have diffraction data confirmed the phase purity of synthetic samples. Here, we report single crystal synthesis, transport and optical properties of \aas\ and a compare  computed electronic structures of the experimental and computationally-relaxed phases.

\section{Methods}

%single crystal characterization
Single crystals of \aas\ with grains larger than 2mm were synthesized. Ag (99.9\% metals basis), Au (99.98\% metals basis), and Se (99.9999\% metals basis) powders were mixed in a stoichiometric ratio inside an Ar filled glove box and sealed in an evacuated silica tube with 7~mm inner diameter. The ampule was placed in a muffle furnace and heated to 750$^{\circ}$C at 1.5$^{\circ}$C/min, annealed for 24 h, cooled to 700$^{\circ}$C at 0.03$^{\circ}$/min, then cooled to room temperature over 24 h. The resulting product was a solid ingot with metallic luster. A portion of the ingot was crushed and analyzed using powder X-ray diffraction (PXRD) on a Bruker D8 diffractometer with a Cu X-ray source in reflection geometry. The crystal structure was refined using the Rietveld method with GSAS-II\cite{toby_gsas-ii_2013}. 

Ingots of \aas were manually polished with SiC grinding papers and alumina slurries down to 0.3 $\mu$m.
%on the Buehler MetaServ 250 polisher. 
Electron backscatter diffraction (EBSD) was performed using a Thermo Scios 2 DualBeam scanning electron microscope (SEM). 
We found that \aas\ surface oxidizes within minutes of exposure in air, which made it difficult to obtain clear Kikuchi patterns. To avoid this issue, mechanically polished samples were milled with 7keV Ar$^+$ ions (Gatan PECS II) immediately before transferring into the SEM chamber. 
Resistivity measurements were carried out using the 2-point contact probe method in a Quantum Design PPMS DynaCool. A rod of \aas\ (1.5 $\times$ 2.8 $\times$ 0.7 mm) was mounted 
%on an ETO puck 
with Kapton tape and 2 gold wire contacts were made by silver epoxy.

Spectroscopic ellipsometry measurements were performed using M-2000 and IR-VASE ellipsometers (J. A. Woollam Co.). Optical conductivity was measured in the UV-visible (0.74-6~eV) and mid-near IR (70-560 meV) energy ranges. Both ellipsometeers were calibrated with a SiO$_2$ (25 nm)/Si wafer prior to scanning. The experimental $\Psi$ and $\Delta$ parameters were obtained at $70^\circ$ incident angle, then converted into the real and imaginary parts of the optical constants to satisfy the Kramers-Kronig relation. The optical conductivity determined at $60^\circ$ and $80^\circ$ did not change significantly, as expected for a cubic material without optical anisotropy. 

First-principles density functional theory (DFT) simulations were performed using Quantum Espresso (QE) \cite{QE-2017} and the Vienna Ab Initio Simulation Package (VASP) \cite{VASP1,VASP2} with  projector-augmented wave pseudopotentials. We used a variety of approximations: (1) The local density approximation (LDA), (2) the generalized gradient approximation with the Perdew-Burke-Ernzerhof parameterization (PBE) \cite{perdew1996generalized}, (3) the modified Becke-Johnson (mBJ) method \cite{mBJ} and (4) the Heyd–Scuseria–Ernzerhof (HSE) approximation with the HSE06 parametrization \cite{HSE06}. Structural relaxations were performed with QE using PBE and a plane-wave basis with a kinetic energy cutoff of 100~Ry and a 7$\times$7$\times$7 $k$-mesh, which is stopped when pressures are below 0.1~kbar. Band calculations were performed using VASP in the presence of spin-orbit coupling, an energy cutoff of 500~eV and a 10$\times$10$\times$10 (5$\times$5$\times$5) $k$-mesh for LDA, PBE and mBJ (HSE06).

\section{Results and Discussion}

\subsection{Growth and structure}

\begin{figure}
\centering\includegraphics[width=\columnwidth]{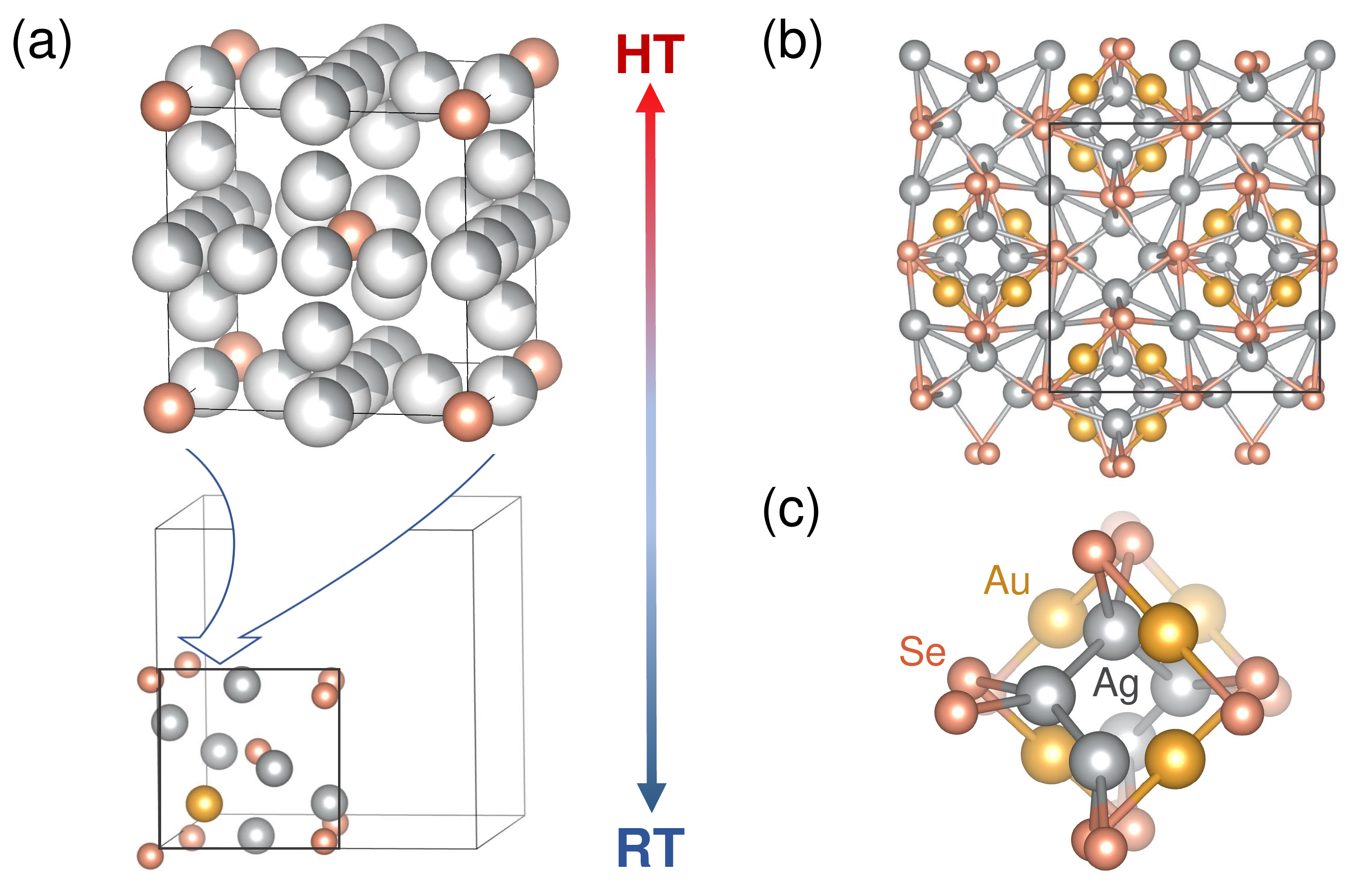} \\
\caption{\label{fig:agause-structure}
The high-temperature $\alpha$-Ag$_2$Se structure with partial Ag occupancy is shown in (a). \aas\ is formed upon cooling by ordering Ag displacements and Au substitution on the  $\alpha$-Ag$_2$Se to form a unit cell that is 8 times the volume. The full unit cell of \aas\ is shown in (b). (c) shows the most apparent chiral $4_1$ chain viewed down $<$100$>$.
%You labeled the atoms in (c), I think it is good
} 
\end{figure}

The structure of \aas\ seems complex, but can be understood as a set of modification from the high-temperature $\alpha$-Ag$_2$Se structure.
Fig.\ref{fig:agause-structure} (a) shows the body-centered-cubic $Im\overline{3}m$  $\alpha$-\agse\ structure in which Se$^{2-}$ ions sit at lattice points,  while Ag$^+$ ions are statistically distributed over 4 positions on and surrounding the face centers. Upon cooling, the metal cations in \aas\ order into one of the available displaced sites, while also distributing the Au$^+$ evenly throughout the cell. This new distorted cell must be doubled in each direction to accommodate the Ag/Au ordering.
The resulting structure of \aas\ (I4$_{1}$32 ($a = 9.96$ \AA\, $Z = 8$) is shown in Fig. \ref{fig:agause-structure}(b), viewed down the $a$ axis. 
The chiral $4_1$ ordering of the cell is best viewed down the $<$100$>$ direction, forming a spiral of Ag and Au in Fig. \ref{fig:agause-structure}(c). 
%Silver atom is coordinated by 4 selenium atoms in a distorded tetrahedron and gold atom is linearly coordinated to 2 selenium atoms.Selenium is 7-coordinated by 6 silver atoms and 1 gold atom.

\begin{figure}
\centering\includegraphics[width=\columnwidth]{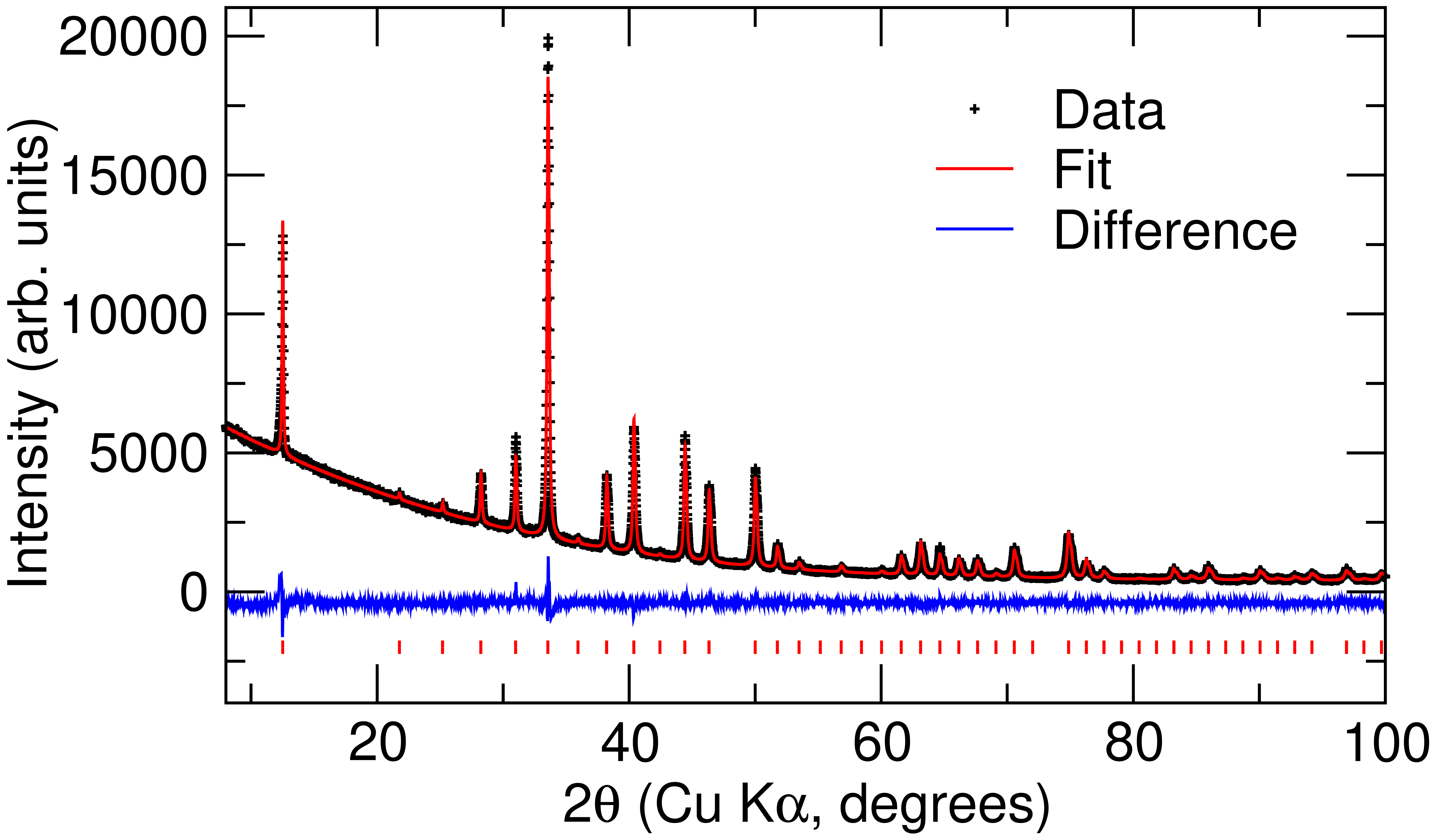} \\
\caption{
Rietveld refinement to X-ray powder diffraction pattern of a crushed ingot of \aas\ measured in reflection geometry.
}
\label{fig:jw25-xrd-gsas}
\end{figure}

\begin{figure}
\centering\includegraphics[width=\columnwidth]{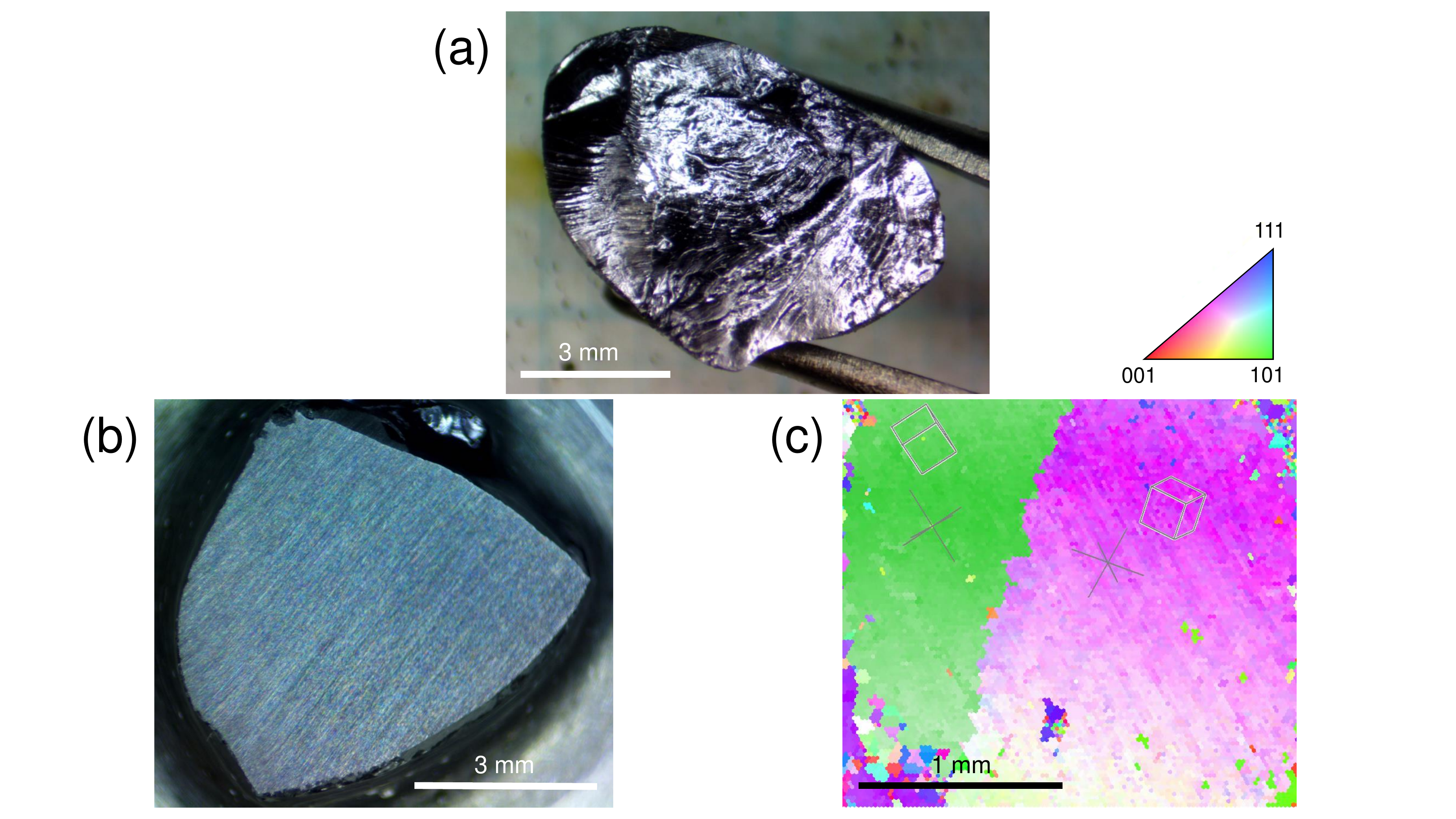} \\
\caption{\label{fig:ebsd}
Fractured and polished (1200 grit SiC) surfaces of \aas\ are shown in (a) and (b), respectively. EBSD grain mapping in (c) indicates grain sizes larger than 2~mm. The green and purple regions have approximate $<$101$>$ and $<$112$>$ directions normal to the sample surface. 
}
\end{figure}

Crushed ingots of \aas\ gave PXRD data shown in Fig. \ref{fig:jw25-xrd-gsas}, which clearly indicate phase purity and the 110 peak at $2\theta = 12.6^\circ$ that indicates the doubled $I4_132$ ordering described in Fig. \ref{fig:agause-structure}. As expected for a cubic material and first noted by Johan,\cite{johan_fischesserite_1971} \aas\ does not cleave leaving large surfaces, with a cross section of the fractured ingot shown in Fig. \ref{fig:ebsd}(a). The material forms large grains after cooling from the melt, so determining the grain structure from polished surfaces can be a challenge because there may be only one or few grains in cross section, such as the polished face in Fig. \ref{fig:ebsd}(b). Therefore, EBSD is necessary to confirm the orientation of various grains in the face. A typical cross section under EBSD mapping is shown in Fig. \ref{fig:ebsd}(c).

\subsection{Electronic properties}

\begin{figure}
\centering\includegraphics[width=\columnwidth]{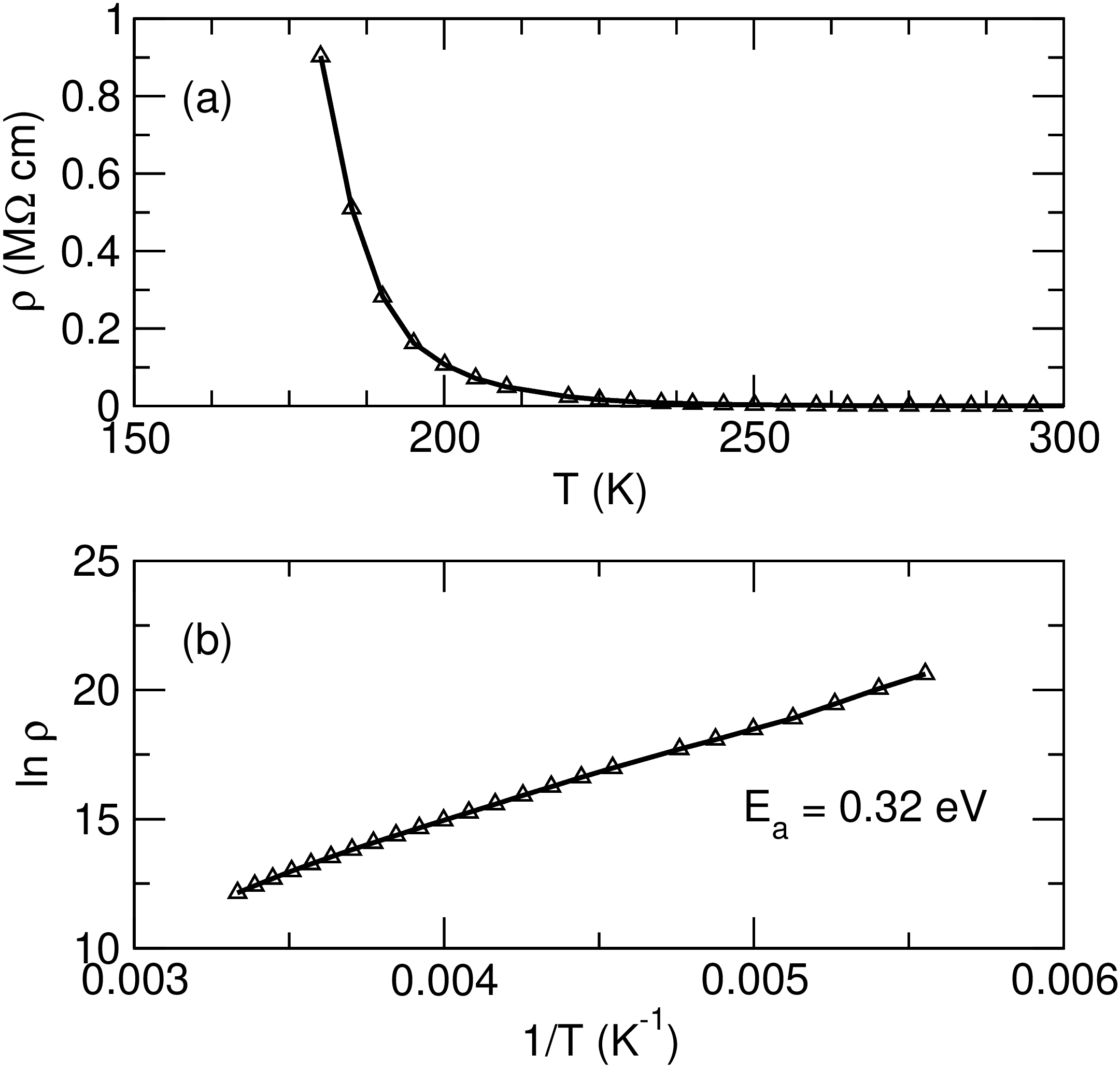} \\
\caption{\label{fig:resistivity}
Two-point electrical resistivity (a) of \aas\ shows semiconducting behavior, while a linear fit to the Arrhenius plot in (b) gives an  activation energy $E_a=0.32$~eV.
} 
\end{figure}

The electrical resistivity and corresponding Arrhenius plot of \aas\ are shown in Fig.\ref{fig:resistivity}. 
We find that \aas\ is highly resistive and exhibits a semiconductor-like temperature dependence, with an Arrhenius activation energy $E_a = 0.32$~eV, which is close to the previously reported value of 0.5~eV from Wiegers.\cite{wiegers_electronic_1976} Given the different microstructures (crystal versus powder) and annealing temperatures (750 versus 450$^\circ$C) between our study and that work, there may be significant differences between the defect populations that scatter carriers in the two resistivity measurements, but the activation energies are the same order of magnitude.
The resistivity of \aas\  is orders of magnitude greater than that of \aat\ in the same structure type.\cite{young_thermoelectric_2000}

\begin{figure}
\centering\includegraphics[width=\columnwidth]{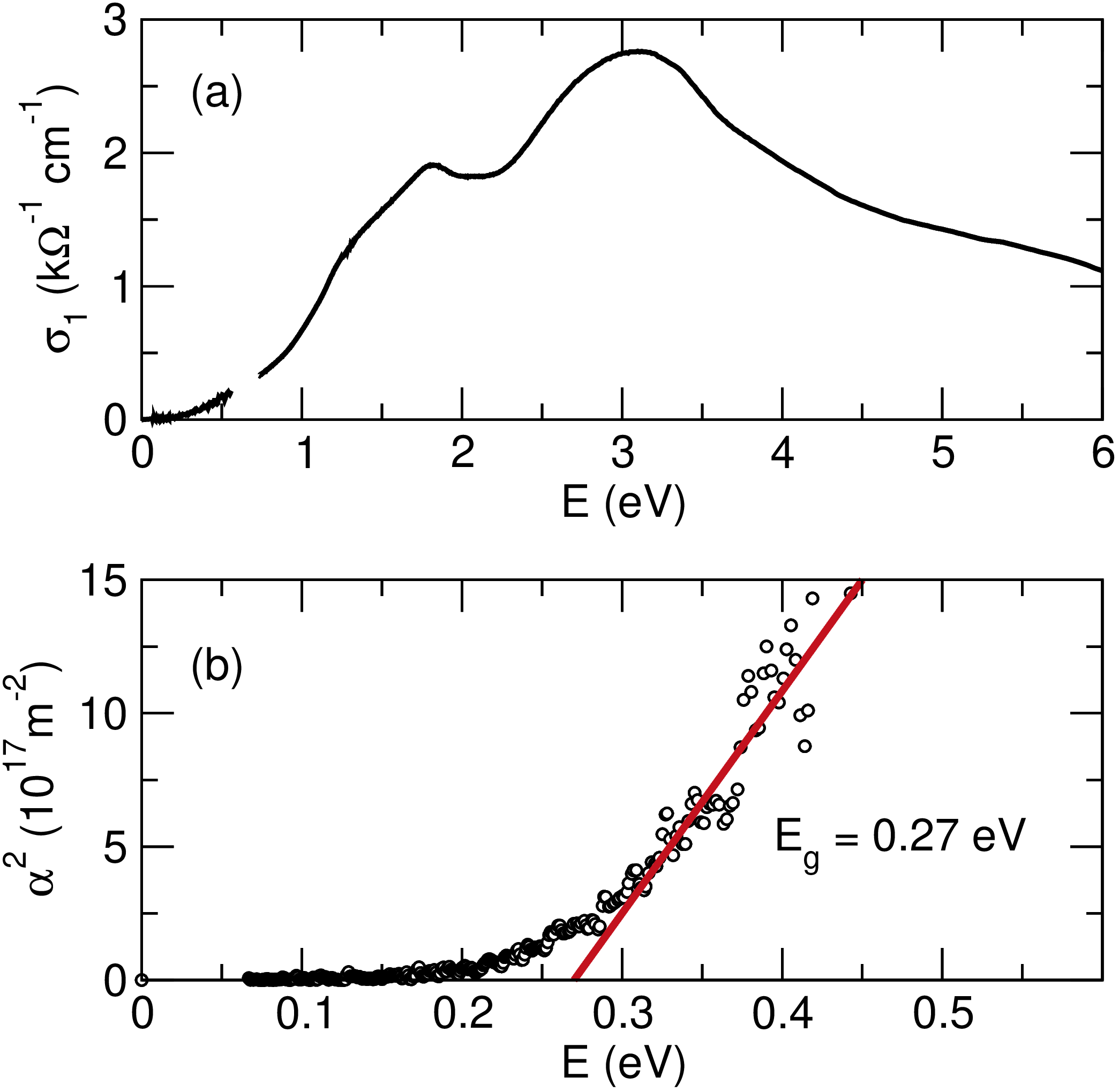} \\
\caption{
\label{fig:ellipsometry} Optical conductivity $\sigma_1$ (a) and the square of optical absorption coefficient $\alpha$ (b) for \aas\ are determined by spectroscopic ellipsometry. In (b), the direct band gap edge $E_g$ of 0.27 ($\pm$0.05) eV shown, which is estimated using the linear relation of $\alpha^2$ and ($E-E_g$) of semiconductors \cite{fox2001}.
}
\end{figure}
% Soyeun comment
% In (b), it would be nice to move the Eg=0.27eV label above the band edge and add a vertical arrow in between. Same arrow (w/o label) can be added in (a).
% (before SYK change) Missing data points between 0.56 and 0.74~eV in (a) are due to a change of light source from the mid-near-IR to UV-visible ranges. => (Actually total 3 lamps were used.) I think this part is already covered in the method and can be omitted.
% Optics people usually plot the inverse of resistivity (1/rho) together at E = 0. For example, see Fig.1d and caption in "https://www.nature.com/articles/s41567-020-0955-0"

In order to investigate the electronic band structure of \aas, we performed spectroscopic ellipsometry measurements. Fig. \ref{fig:ellipsometry}(a) shows the experimental optical conductivity $\sigma_1$ of \aas\ measured at room temperature. No sign of free carrier response (intraband transition) was captured in the $\sigma_1$ down to the lowest frequency measured, consistent with the measured high resistivity and finite gap predicted from previous calculations. \cite{fang_ab_2002,faizan_elastic_2016,faizan_carrier_2017}
%Optical measurements are closely related to band structure and we can draw certain analogies about \aas\. 
% Direct band gap extraction
The band gap edge of \aas\ was extracted from the squared optical absorption coefficient $\alpha^2$ as in Fig. \ref{fig:ellipsometry}(b). From the Fermi's golden rule and the joint density of states for parabolic bands, a frequency dependence of absorption near a direct band gap ($E_\text{G}$) is represented as \cite{fox2001}:
\begin{equation}
   \alpha \propto \sqrt{\omega -E_\text{G}}\label{eq:bandgap},
\end{equation}
for $\omega > E_\text{G} $. Previous band structure studies of \aas, and our own, support parabolic curvature of the band edges of the direct gap at the Brillouin zone center  $\Gamma$.\cite{fang_ab_2002,wiegers_electronic_1976,faizan_elastic_2016} 
Using the above relation, the direct band gap of \aas\ is estimated to be 0.27~eV, as illustrated in Fig. \ref{fig:ellipsometry}(b). Uncertainties in the exact onset of the rise in $alpha^2$ and the imperfect geometry of the resistivity sample measured in Fig. \ref{fig:resistivity} lead to a rough agreement of both measurements to $\sim 0.3$~eV.

Above an excitation energy of 0.27~eV the optical conductivity gradually increases and shows two clear peaks near 1.8 and 3.2~eV. These peaks arise from higher energy interband transitions, i.e. optical excitation between the valence and conduction bands that are further away from the Fermi energy. Note that there may be additional interband transitions due to the observed kink-like features in $\sigma_1$ at 1.3~eV, 4~eV and 5.5~eV.

\begin{figure}
\centering\includegraphics[width=\columnwidth]{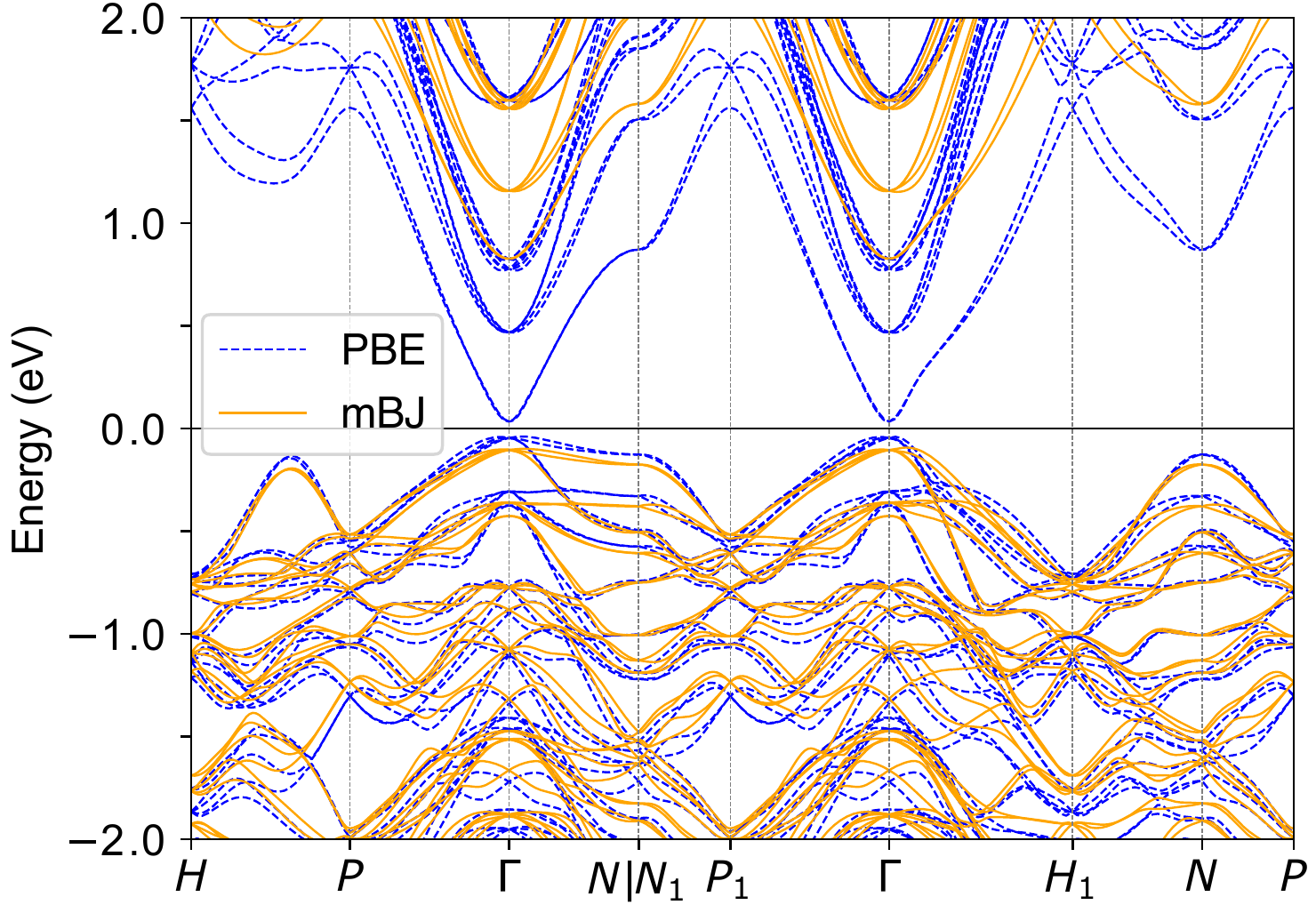} \\
\caption{\label{fig:DFT-bands}
DFT-calculated electronic band structures of \aas\ as obtained from PBE and mBJ functionals. The general features of the band structures are equivalent for both functionals, but the direct band gap is predicted to be 71 meV for PBE and 907 meV for the hybrid mBJ functional. 
}
\end{figure}

\begin{figure}
\centering\includegraphics[width=0.9\columnwidth]{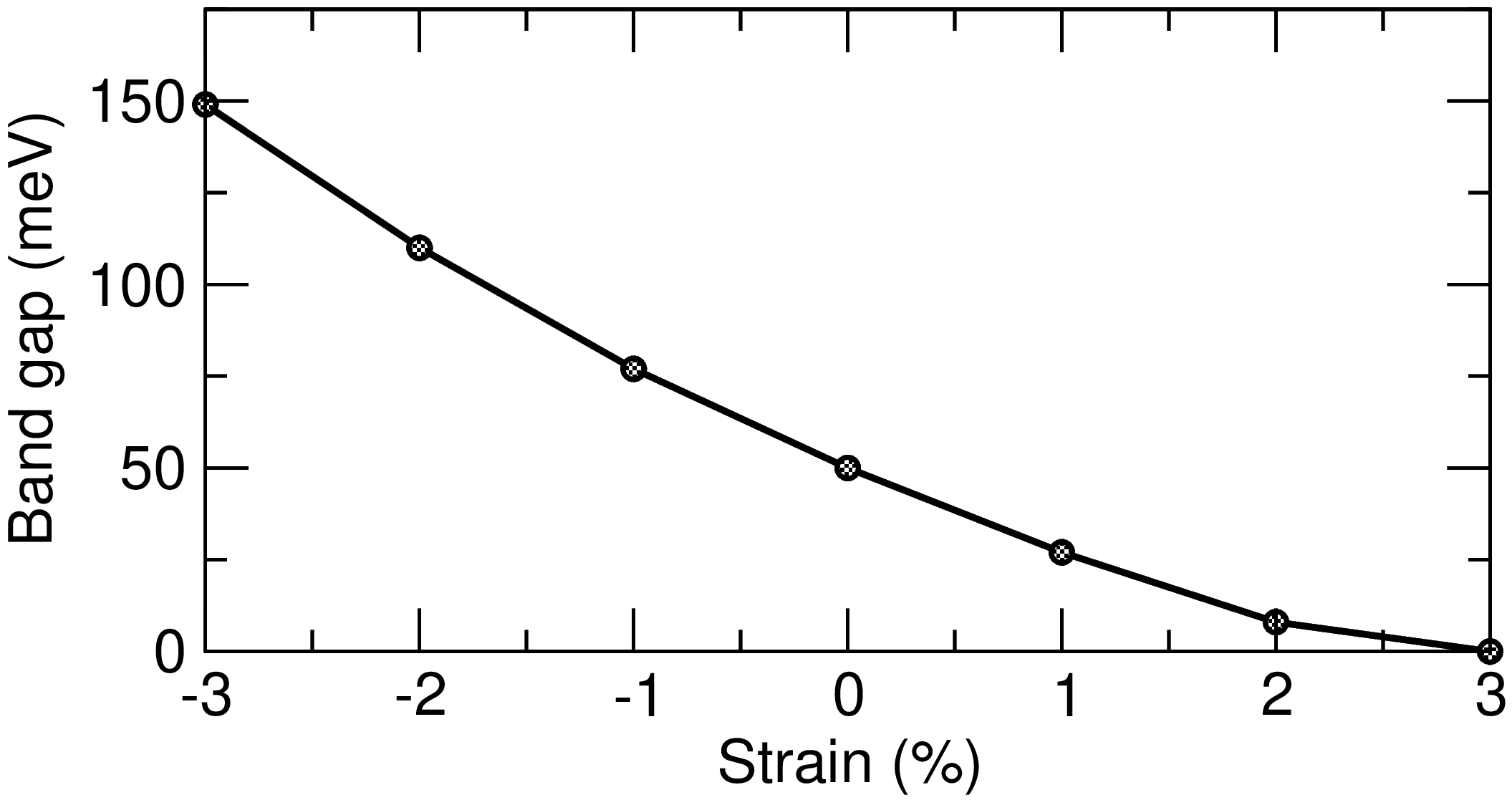} \\
\caption{\label{fig:DFT-strain}
 The DFT-PBE band gap of the relaxed \aas\ structure decreases with tensile strain, leading to band overlap around 3\% strain.
}
\end{figure}

The DFT-calculated band structures of \aas\ from PBE and mBJ functionals are shown in Fig. \ref{fig:DFT-bands}. Although the gaps differ by about 1~eV, the shape and features are consistent for both functionals. A direct gap is observed at $\Gamma$ and bands are symmetric around $P_1$.
Gaps of \aas\ were calculated using multiple approximations for both the experimentally-refined and DFT-PBE relaxed structures. 
The obtained gaps are listed in Table \ref{tab:DFT-gaps}.
The gaps span a wide range, from fully overlapped metallic bands to 931 meV. The differences between experimentally-refined and DFT-relaxed structures for a given functional are typically no larger than 30~meV, whereas the differences between LDA/PBE and the hybrid functionals mBJ/HSE06 for a given structure is much larger, over 800~meV. It is important to note that although the 30~meV differences are comparatively small, subtle differences in the atomic positions will have a large effect on transport if the band gap is of the same order of magnitude (tens of meV). 
These small effects are also evident in shifts of the bands due to increased overlap with compressive strain.
The PBE band gaps of the \aas\ DFT-PBE-relaxed structures with varying lattice strain are plotted in Fig. \ref{fig:DFT-strain}, showing an approach to metallicity as the tensile strain reaches 3\%, which is a large value of strain to be reached by mechanical compression, but may be attainable by chemical substitution of the Se$^{2-}$ anion in particular.

\begin{table}[]
\begin{tabular}{c|c|c}
%\cline{2-3}
                            & Experimental structure & DFT-PBE structure \\ \hline
LDA   & No gap                 & No gap                \\ %\hline
PBE   & 71 meV                 & 50 meV                \\ %\hline
mBJ   & 931 meV                & 928 meV               \\ %\hline
HSE06 & 907 meV                & 872 meV               \\ %\hline
\end{tabular}
\caption{\label{tab:DFT-gaps} DFT-calculated band gaps of \aas\ for experimentally-refined structures and DFT-PBE-relaxed structures. Four pairs of ab-initio calculations using the LDA, PBE, mBJ and HSE06 approximations for the exchange correlation energy are used.}
\end{table}

The noticeably larger band gap found by the hybrid mBJ and HSE functionals points to additional disorder that may lead to defect states that dominate the electronic properties of \aas.
Percent-level antisite defects in the Ag/Au ordering that occurs at the 270$^\circ$C order-disorder transition would be challenging to quantify, and would likely remain coherent with the large-grained microstructure. Additional single-crystal diffraction may shed light on this possibility. Additionally, more detailed investigation of the optical properties, including angle-resolved photoemission spectroscopy, should correlate larger features of the conduction band with first-principles results, rather than states exclusively near band edges. Understanding these details for the entire class of fischesserite-type phases will reveal their potential for nonlinear optics and topological responses.

\section{Conclusions}
We have established that large single crystals of \aas\ can be grown by slow-cooling from the melt.
Ellipsometry and resistivity measurements point to an activation energy and optical band gap around 0.3~eV, which is between the PBE-estimated band gap of 0.07~eV (which should be an underestimation) and the mBJ and HSE06 band gaps of about 0.9~eV, which we expect to be accurately predicted. 
Given the sensitivity of the band gap to strain, and the necessary order-disorder transition that \aas\ experiences on cooling, it follows that subtle tuning of the defect population, surface treatment, and stoichiometry of \aas\ and the related fischesserite-petzite family of materials can lead to intentional tuning of the band gap. It remains to be seen whether a band inversion can be obtained upon application of pressure of chemical substitution with the telluride.

\section{Acknowledgments}

This study was supported by the Center for Quantum Sensing and Quantum Materials, an Energy Frontier Research Center funded by the U.\ S.\ Department of Energy, Office of Science, Basic Energy Sciences under Award DE-SC0021238.
The authors acknowledge the use of facilities and instrumentation at the Materials Research Laboratory Central Research Facilities, University of Illinois, partially supported by NSF through the University of Illinois Materials Research Science and Engineering Center DMR-1720633.
MGV, IE and MGA acknowledge the Spanish Ministerio de Ciencia e Innovaci\'{o}n
(grant number PID2019109905GB-C21) and Programa Red Guipuzcoana de Ciencia,
Tecnolog\'{i}a e Innovaci\'{o}n 2021 No. 2021-CIEN-000070-01 Gipuzkoa Next. MGV thanks support from the DeutscheForschungsgemeinschaft  (DFG,  German  Research  Foundation)  GA  3314/1-1  –  FOR5249  (QUAST).
SB acknowledges support through the Early Postdoc Mobility Fellowship from the Swiss National Science Foundation (Grant number P2EZP2 191885).
% SNU acks
BL, JSS, and TWN were supported by the Institute for Basic Science (IBS) in Korea (Grant No. IBS-R009-D1).

\bibliography{agause}

\end{document}